# Readout Optical System of Sapphire Disks intended for Long-Term Data Storage

V.V.Petrov[1], V.P. Semynozhenko[2], V.M. Puzikov[2], A.A. Kryuchyn[1],

A.S. Lapchuk[1], S.M. Shanoilo[1], I.V.Kosyak[1], Yu. O. Borodin[1], I.V. Gorbov[1], Ye.M. Morozov[1,*]

[1] Institute for Information Recording, National Academy of Science of Ukraine,
2 Shpak Str., 03113, Kyiv, Ukraine

[2] STC "Institute for Single Crystals", National Academy of Science of Ukraine,
60 Lenin Ave., 61001, Kharkiv, Ukraine

The development of long-term data storage technology is one of the urging problems of our time. This paper presents the results of implementation of technical solution for long-term data storage technology proposed a few years ago on the basis of single crystal sapphire. It is shown that the problem of reading data through a substrate of negative single crystal sapphire can be solved by using for reading a special optical system with a plate of positive single crystal quartz. The experimental results confirm the efficiency of the proposed method of compensation.

## 1. Introduction

One of the major problems facing mankind is the preserving and transmitting obtained knowledge to the succeeding generations. Its solution is not possible without creating a media that would provide long-term data storage. A long shelf life of data storage is generally understood as the data retention that greatly exceeding the shelf life of information recording on the best acid-free paper having the shelf life of about 200 years. However, the real time of data storage in modern storage devices ranges from 3 to 30 years.

The optical discs have a great potential for long-term archival storage. However, the modern optical discs can not provide the required level of reliability and data retention time due to the low stability of the polycarbonate substrate [1], and not sufficiently high its adhesion to the metal layer. The variety of technical solutions is proposed for technology of long-term storage which is based on using highly stable materials for both substrate and recording mediums.

* Corresponding author: more_geniy@mail.ru

First designed and manufactured optical discs had a substrate of silica glass [2]. Their data retention time could not exceed of 20-30 years because of time degradation of recording layer based on chalcogenide glasses [3].

The cermets are used as recording medium in long-term optical M-discs of write-once format with shelf life of 150-160 years [4]. A borosilicate glass is proposed as medium of the disc substrate as well as refractory metals such as tungsten and platinum are proposed to use for recording layer. An embossed microstructure on the surface is prepared by ion etching [5]. It is proposed to seal tungsten alloy disc by silicon nitride [6] and to read the information from platinum layer through sapphire windows [7]. The quartz is also proposed as substrate of optical disc for long-term data storage [8]. The establishment of such data storage focuses on long-term and reliable data storage, rather than getting the limit values of density and large amounts of data storage [5, 6, 8, 9].

## 2. Foundation for the choice of optical disc substrate medium and the peculiarities of optical scheme of recording system

Materials for the manufacture of optical discs having long-term data storage must be chemically, thermally and mechanically stable. The main elements determining parameters of the optical disc are a transparent layer (substrate) through which the information is reading and recording layer on which information is recorded. Disc substrate layer is an external unprotected and it defines the types of materials and technologies that can be applied in the deposition of the recording layer, so the choice of the substrate medium is the key element in development of technology of long-term data storage.

One of the main causes of data loss is the appearance on the substrate surface roughness and scratches. Therefore, the optical disc substrate must be formed from the hard as possible material. The reliability to multiple data retrieving requires that the substrate medium was resistant to light and ultraviolet radiation. For the stability of the optical disk to large temperature changes and the possibility of using refractory alloys as recording layers, the substrate material should be a refractory one. Another important factor of the reliability is high thermal conductivity of the substrate as small thermal conductivity can cause localized heating and as a result, to permanent deformation of the disc. Furthermore, when using a refractory metal as a material of the recording layer, the linear coefficient of thermal expansion of disc substrate must not significantly differ from values for the metals.

The similar requirements for physical and technical characteristics are imposed on optical materials that are used for optical sensors and observation windows in the chemical, nuclear and space industries. Therefore, as a substrate for non-volatile storage can be used the materials that are applied for the optical sensing and observation windows in these industries. It should be noted that there is one significant limitation on the optical characteristics of the substrate material, namely, the optical anisotropy of the optical disc substrate should not go beyond the uniaxial birefringence because otherwise the compensation of aberrations being caused by the anisotropy of the substrate is a problematic task. Table 1 summarizes both the optical and physical characteristics of the materials that can be used for optical disc data retention.

| Parameters \ Material | Sapphire ($Al_2O_3$) | Quartz ($SiO_2$) | Fused quartz ($SiO_2$) | Yttrium aluminum garnet ($Y_3Al_5O_{12}$) | Magnesium aluminate ($MgAl_2O_4$) | Diamond (C) |
|---|---|---|---|---|---|---|
| State of matter | Crystalline | Crystalline | Amorphous | Crystalline | Crystalline | Crystalline |
| Optical type of crystal | Negative | Positive | – | – | – | – |
| Mohs hardness | 9 | 7 | 5,3 – 6,5 | 8,5 | 8 | 10 |
| Fusing temperature (K) | 2300 | 1960 | 1350 (softening) | 2210 | 2400 | 1100 (burn down) |
| Thermal conductivity coefficient $k$ (W/(mK)) | ~34 | 3 | 1,3 | 14 | 15 | 3200 |
| Coefficient of linear thermal expansion $\alpha*10^{-6}$ (1/K) | 5,6 | 0,55 | 0,55 | 8,0 | 7,5 | 1,0 |
| Chemoresistance | 1* | 2* | 3* | 1* | 3* | 1* |
| Resistance to UV radiation | Not degraded | Not degraded | Degraded | Not degraded | Not degraded | Not degraded |

Table 1. Physical and chemical properties of materials suitable for the manufacture of optical discs for long-term data storage:

1* - soluble in concentrated solutions of fluoride salts (such as, $BaF_2$, $MgF_2$, $PbF_2$, and oxides) at very high temperatures (> 1000 ° C);

2* - soluble in alkaline aqueous solutions at high temperatures (> 300 ° C) and solution of hydrofluoric acid;

3* - soluble in aqueous alkaline solutions and solution of hydrofluoric acid;

From the data in Table 1 is clear that sapphire, quartz, yttrium aluminum garnet and magnesium aluminates have physical characteristics required for long-term data storage and therefore can be used as medium fore substrates of the long-term optical disc.

Sapphire is the hardest material being made of cheap components, and its technology for growing single crystals is simple and affordable process. The disadvantage of sapphire is its sufficient optical anisotropy (birefringence) and as a consequence, the complexity of the optical scheme for reading information in the sense of the aberrations compensation caused by the anisotropy.

Yttrium aluminum garnet and magnesium aluminate have got a cubic lattice, and, consequently, they are optically isotropic materials (not exhibit birefringence). Therefore, reading out through the substrate made of these materials does not require any additional methods of the aberrations compensations caused by the substrate anisotropy. Their disadvantages are smaller than sapphire hardness and the production of optical discs with a diameter of either 80 mm or 120 mm is a very labor-intensive and costly.

Diamond has several unique physical characteristics. However, there is no technology of single-crystal diamond with dimensions required for optical recording and the actual production of mono crystalline diamond is extremely expensive. Furthermore, in ambient atmosphere at a temperature of 1100 °C and more the diamond burns.

Single crystalline quartz is considerably inferior to sapphire in hardness and is a uniaxial crystal. Fused silica has no optical anisotropy, but it has a considerably lower compared with the other materials listed in Table 1, hardness and is sensitive to ultraviolet (darkens).

From the above analysis of physical and technical properties of the materials, it is clear that sapphire is one of the best materials for the manufacture of substrates of optical disc for long-term storage of information. In favor of the sapphire is the fact that among of the crystals, the production of which is mastered in significant quantities and already having a comparatively low cost of manufacturing, the leader is the sapphire ($Al_2O_3$). The analysis of existing highly stable uniaxial single crystal materials has demonstrated that the best material for the optical disc substrate is leucosapphire [9, 10], on the internal surface of which the information micro relief structure is formed [5, 9, 10] since it has a high chemical stability, its wear resistance is 8 times greater than that of steel, it is thermally stable up to 1600 $^0$C and it is optically transparent in the range from 0.17 microns to 5.5 microns [11].

The sapphire birefringence is cause of aberrational distortions of the scanning laser beam as it passes through the optical disc. This is posed by the difference in optical path of light rays with different polarization that makes impossible to use the standard optical system for reproducing the information from the optical disc with a birefringent substrate. The orientation of the sapphire single crystal axis 0001 parallel to the optical axis of the optical head will results in a constant magnitude of aberrations during rotation of the optical disc. However the magnitude of aberrations is so large that the resolution of the optical system drops 4.3 times for the case of numerical aperture of 0.4 and it becomes more significant with an increase of numerical aperture. Accordingly, the use of the birefringent single crystal materials as substrates for optical discs requires a change of standard optical reading out system so as to compensate the aberrations created by rays with different polarization. This minimization of the polarization effect can only be applied to those optical discs manufactured from uniaxial birefringent materials with the optical axis orthogonal to the disc surface.

The sapphire substrate on which information is recorded in the form of a micro relief structure, and the data are read through the sapphire substrate can be a key element of technology for optical data storage with high reliability and long shelf life [9, 10]. Thickness of the substrate may be 0.4-1.2 mm, which provides sufficient mechanical strength to the carrier. Sapphire substrates have significant birefringence ($n_0$ = 1.78038 and $n_e$ = 1.77206, $\lambda$ = 442 nm), which leads to significant aberrations when focusing the laser emission through the substrate of the carrier.

A superposition of astigmatism and spherical aberrations of different orders arise from different optical path (spherical front phase distortions) of s- and p- polarized peripheral rays of focused laser beam propagating through sapphire substrate. In particular, spherical front phase distortions lead to the fact that radiation with orthogonal polarizations is focused in different places, the distance between which $\Delta F$ is defined as follows:

$$\Delta F = 2h\Delta n / n_0, \qquad (1)$$

where $\Delta n$ is the difference between the refractive indices of the ordinary ($n_0$) and extraordinary ($n_e$) beam (for sapphire $\Delta n = 8 \cdot 10^{-3}$), $h$ - thickness of the substrate.

When focusing through the sapphire substrate with thickness of 1 mm, the distance between the spots is several times larger than the depth of focus. This makes it impossible to reliably reproduce the recorded data. There is a possibility to compensate the aberrations caused by the anisotropy of the sapphire substrate using an additional compensating plate with a positive uniaxial material (inverse to the sapphire). The study of available transparent uniaxial optical crystals showed that the best material for compensating aberrations of the sapphire substrate is quartz that has $n_0$ = 1.5443, $n_e$ = 1.5534. Conditions for obtaining the optical system with minimal residual aberrations can be written as:

$$H_{kvr} / H_{spf} = 0{,}68; \quad H_{kvr} + H_{spf} = H, \qquad (2)$$

where $H_{kvr}$ and $H_{spf}$ are the quartz plate and the sapphire substrate thicknesses, respectively. First condition follows from astigmatism aberration compensation and second – from spherical aberration compensation. Table 2 shows the thickness of the sapphire substrate and compensation quartz plate for optical discs in various formats.

| Media type | The thickness of sapphire substrate, (mm) | The thickness of quartz substrate, (mm) | Uncompensated aberration $\Delta\Phi/\Phi_0$, 100% |
|---|---|---|---|
| CD | 0,714 | 0,486 | 1% |
| DVD | 0,357 | 0,243 | 1,6% |
| CD (for reading at λ = 400 nm) | 0,714 | 0,486 | 0,3% |

Table 2. Thickness of the sapphire substrate and the quartz plate compensation for optical discs of different formats

Hence it is possible to carry out reproduction of data by reading through a sapphire substrate by using special focusing optical system comprising a quartz plate. Such carriers can be used for long-term data storage due to the high thermodynamic stability of sapphire. Schematic representation of the method of compensation of aberrations is shown in Fig. 1.

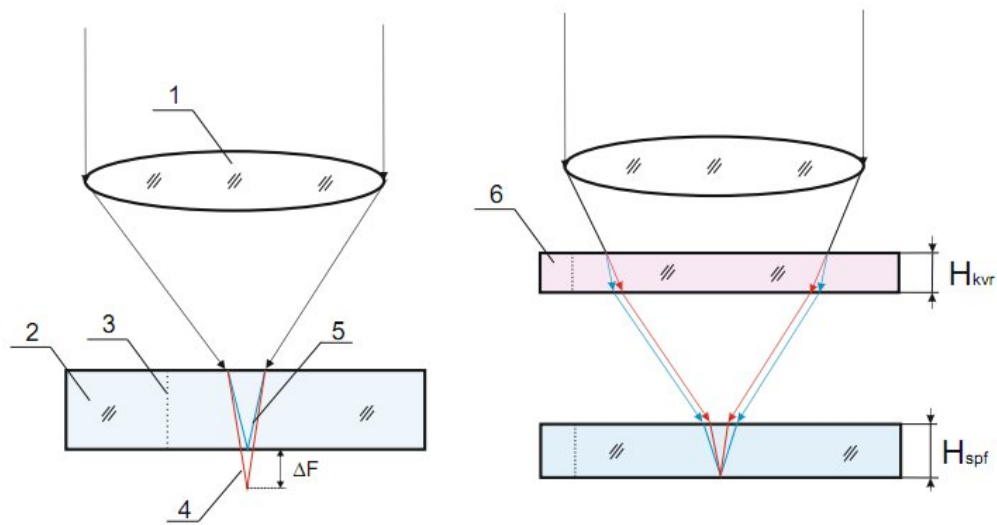

Figure 1. Schematic representation of the propagation of the ordinary and extraordinary rays:
1 - lens, 2 - sapphire substrate, 3 - optical axis, 4 - extraordinary ray, 5 - ordinary ray, 6 - quartz plate.

The study of existing transparent uniaxial optical crystals demonstrated that the best material for compensation of aberrations in the sapphire substrate is quartz, which has the following refractive indices: $n_0 = 1,5443$, $n_e = 1,5534$.

## 3. Experimental results

The information (in a standard CD-ROM format) was recorded (at 2X speed) on photoresist layer with thickness of 150 nm deposit on sapphire disc having a thickness of 0.7 mm and a diameter of 80 mm by audio recording laser station established at the Institute for Information Recording of NAS of Ukraine. A single-mode semiconductor laser QLD-405-100S (405 nm, 100 mW) was used as a source of coherent light. Power of the laser beam at the output of the lens was 3 mW. Etching time of the photoresist layer in an alkaline etchant consisted 7 seconds. A modified optical ROM drive Plextor PX-891 SA was used for the optical system test.

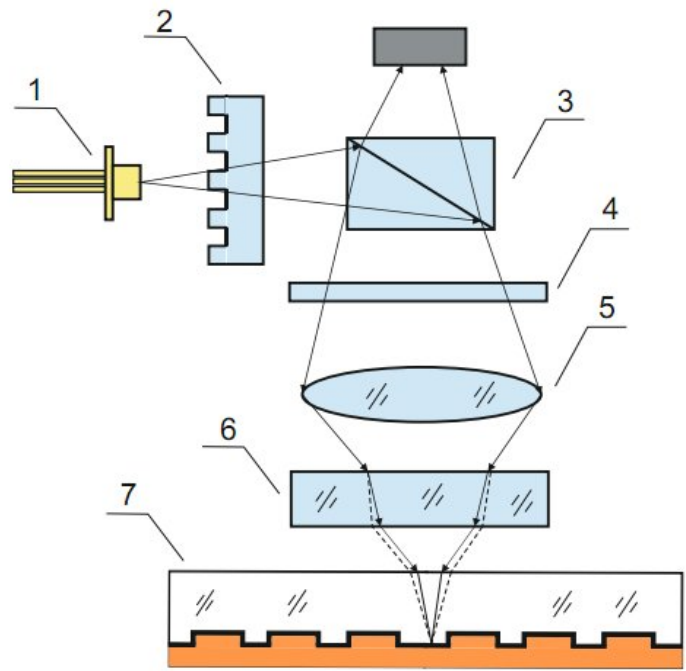

Figure 2. The optical system for reading data from sapphire disk.

Fig. 2 shows the optical scheme for data readout from sapphire optical disc [11]. During data reproduction, the laser beam from a laser diode (1) passed through the diffraction grating (2), a quarter-wave plate (4), and the beam splitter cube (3) and is directed onto the focusing objective lens (5). The focusing lens focuses the laser beam through a compensating plate (6) and a substrate carrier (7) on the microstructure relief of information carrier. The standard reading device modification is consisted in fixing a compensating quartz plate (with vertical orientation of optical axis) having a size 5x5x0.5 mm to the lens by gluing method.

The presence of a compensating plate made of uniaxial single crystal material leads to the compensation of optical path difference between s- and p- polarized rays and focusing them in one focal plane. The results of test pointed out that the quartz compensating plate allows a robust data retrieving in the case of reading data through sapphire substrate by a standard optical drive. Figure 3 demonstrates the result of test obtaining by Nero CD-DVD Speed program, from which one can see that the disc does not have uncorrectable errors C2, but there are small numbers of correctable errors C1.

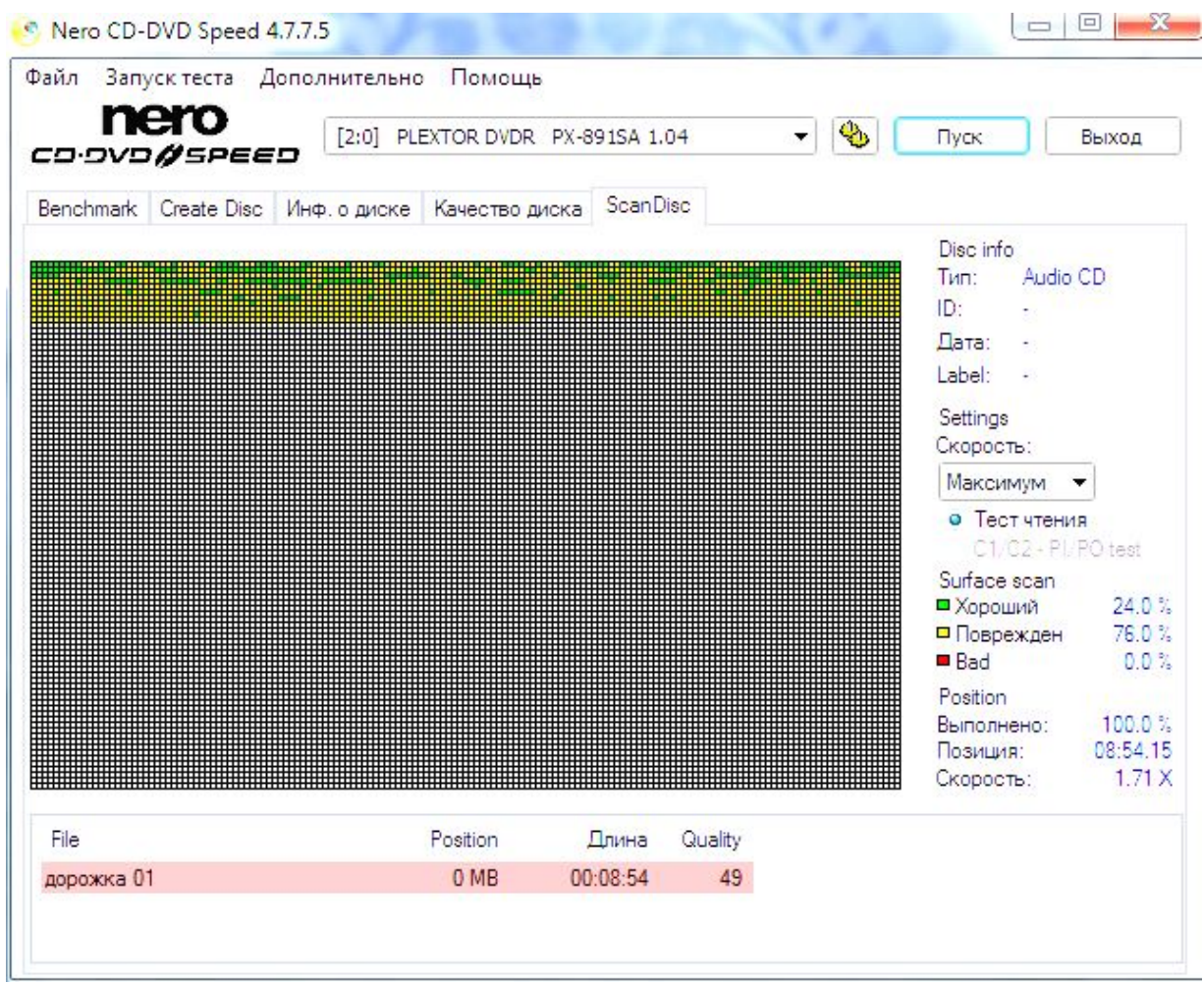

Figure 3. The results of the test of retrieving data through sapphire substrate reading out by modified standard reading device. The results obtained by Nero CD-DVD Speed program.

The experiment has pointed out the possibility of full compensation of polarization stipulated wave front distortion of the sapphire substrate. The individual method of sapphire substrates manufacture allows obtain more accurate substrate thickness than in injection molding process of polycarbonate optical discs. The transparency of sapphire in the ultraviolet range allows using shorter wave radiation for recording and reproducing information.

Almost all possible types of optical discs can implement on the basis of sapphire substrates:
- optical disc with thermally treated and metalized photoresist layer;
- CD-ROM with a high-reflective coating;
- the magneto-optical and CD-RW;
- CD-R disk.

CD-ROM with a high-reflective coating is the primary format for the implementation of optical discs with a shelf life of optical discs within the tens of thousands of years or even more. The main limiting factor for the shelf life can be adhesion of reflective coating. The most long-term data storage can be hopefully got on the hermetic CD-ROM drives without a reflective coating. Realization of the magneto-optical and CD-RW optical drive on a sapphire substrate will allow get more high-optical materials and hence significantly increase a shelf life of data storage. Application of the CD-R technology can also lead to the increase of shelf life of optical disk due to the high chemical resistance of sapphire and the high melting point.

**4. Conclusions**

1. It has been considered the effect of birefringence on the distribution of the focused laser beam through a uniaxial birefringent medium having a vertical orientation of the optical axis. An expression for the calculation of the geometric aberrations of the focused laser beam in single-crystal substrate of the optical disc has been presented.

2. The permissible levels of the anisotropy were calculated for optical discs of different formats allowable thickness discs of different formats and the maximum thickness of sapphire optical discs in which there is no significant signal distortion have been demonstrated.

3. The method of compensating aberrations when reading data from uniaxial birefringent medium was developed. The basic parameters of reading out system for the sapphire disc with a vertical optical axis orientation were calculated. The structural parameters (thickness) of the optical disc sapphire substrate and quartz compensating plate of different formats were presented.

4. An experimental verification of the method of the aberrations compensations which showed that the quality of images obtained through the glass and sapphire plate with compensation, virtually identical has been performed. Thus, the application of the compensating quartz plate further allows read information from the optical sapphire disc for information retention.